\documentstyle[12pt]{article}
\begin{document}
%\baselineskip=1.5\baselineskip
\begin{center}
{\Large Varying $c$ cosmology and Planck value constraints}\\
\bigskip
{\bf D.H. Coule}\\
\bigskip
School of Mathematical Sciences\\
University of Portsmouth, Mercantile House\\
Hampshire Terrace, Portsmouth PO1 2EG
\bigskip
\begin{abstract}

It has been suggested that by increasing the speed of light during the
early universe various cosmological problems of standard  big bang cosmology
can be overcome, without requiring an inflationary phase.
However, we find that as the Planck length and
Planck time are then made  correspondingly smaller, and together with
the need that the universe should not re-enter a Planck epoch, the
higher $c$ models have very limited ability to resolve such problems.
For a constantly decreasing $c$ the universe will quickly becomes quantum
gravitationally dominated as time increases: the opposite to
standard cosmology where quantum behaviour is only ascribed to early
times.

\end{abstract}


\end{center}
\newpage
{\bf 1 Introduction}

It is well known that the standard big bang model (SBB) model has
a number of worrying puzzles, particulary the so-called horizon and
flatness problems, that generally
relate to the fixing of arbitrary
constants. But, as emphasized
 by Zeldovich [1] perhaps the most fundamental and
serious being the fact that the energy
density is much larger than the Planck value $\sim 10^{93}gcm^{-3}$ if
the present universe is run back to when the universe was Planck size.
Equivalently to account for this discrepancy we require that the
size of the universe be already much larger than the Planck
length ($l_{pl}$) for time $\sim$ Planck time $(t_{pl})$. This
mismatch of scales is generally referred to as the Planck problem of SBB
cosmology.
 This is true of any matter
source that obeys the strong-energy condition, so including radiation
or dust sources.

{\bf 1.1 The Planck problem}

Consider just a radiation source with a FRW metric, the Friedmann
equation is,
 \begin{equation}
 H^2+\frac{c^2k}{a^2} = \frac{A}{a^4}\;\;\; (A=\rm{constant})
 \end{equation}
where we set Newton's constant $8G/3=1$, but keep $c$ explicit but
constant throughout section (1).
 The solution of this equation, ignoring the curvature $k$, is simply
 $a=A^{1/4} |t|^{1/2}$, with $t=0$ being the initial singularity.
 We can now highlight the Planck problem that occurs with a big bang model
 with such a matter source.
 If such a model is to account for our present universe then the
 constant $A$ has to be extremely large $\sim 10^{120}$.
  Consider a universe created with Planck
  radius ($\sim10^{-33}$ cm) and Planck
  density ($\sim 10^{93}$gcm$^{-3}$). If such a universe
  expands to its present size greater than $\sim 10^{28}{\rm cm}$
then the density would be of order [1]
\begin{equation}
 \rho \sim 10^{93} \left (10^{28}/10^{-33} \right )^{-4}\;\simeq 10^{-151}
 {\rm gcm}^{-3}
 \end{equation}
  This should be compared to the present energy
  density $\sim 10^{-30}{\rm gcm}^{-3}$ . Even if
  the radiation energy density was immediately converted into dust the
  resulting energy density would still be $\sim 10^{-90}{\rm gcm}^{-3}$.
  To account for this discrepancy
  we require the constant $A$ to be so large $\sim 10^{120}$
  that the energy density is
  now vastly greater (a factor $A$)  than the Planck value for when
  the universe is $\sim$ Planck size . This also forces
  the size of the universe $a\sim A^{1/4}t^{1/2}$
  to be   much bigger than Planck
  size for time $\sim$ Planck time
  ($t_{p}$). This Planck problem  is, as said, in many ways
  the most fundamental problem we first
  need to solve with a cosmological model. Otherwise we will fail
  to understand the enormous size and matter content of our actual universe.
  This problem is present in
  flat $k=0$ universe since for a
  natural value of $A\sim 1$ the size of a radiation dominated universe
  with scale factor $a\sim t^{1/2}$,
  with today's lifetime $10^{60}t_p$ is only $\sim 10^{-33}{\rm cm}*10^{30}
  \sim 10^{-3}{\rm cm}$! . But now including the $A^{1/4}\sim 10^{30}
  $ factor gives a more correct $\sim 10^{27}$ cm value.

  {\bf 1.2 The horizon problem}

The horizon problem occurs because the the particle horizon size,
defined as
\begin{equation}
r=c\int_{0}^{t} \frac {dt}{a(t)}
\end{equation}
is finite, see eg.[2-4]. The horizon proper distance $R$ is
 this quantity $r$
multiplied by the scale factor i.e $R=a*r$. For any strong-energy
satisfying matter source this quantity $R$ grows linearly with
time. But in SSB cosmology the rate of change of the scale factor,
given by $a\sim t^p$ and $1/3<p<1$, grows increasingly rapidly as
$t\rightarrow 0$. The horizon cannot keep pace with the scale
factor `velocity' $\dot{a}\sim 1/t^{1-p}$. But note that this is
only impossible for times below unity $0<t<1$. If the horizon
problem was solved, by some process, at the Planck time $t_{pl}=t= 1$
 it would remain  permanently solved during the ensuing evolution.
For the inflationary value $p>1$ the horizon problem is simply
avoided. One can further understand this by noting that the usual
space-like singularity of the FRW universe becomes a null
singularity when $p>1$- see eg.[5].

 {\bf 1.3 The flatness problem}

 Consider a perfect-fluid equation of state: $p=(\gamma-1)\rho$.
 The Friedmann equation is again
 \begin{equation}
 H^2+\frac{c^2k}{a^2}=\rho
 \end{equation}
 There is also the continuity equation
 \begin{equation}
 \dot{\rho}+3H\gamma\rho=0\;\;\;\Rightarrow\;\;\;\rho=\frac{A}{a^{3\gamma}}
 \end{equation}
 With $A$ the previously introduced constant.
 The density parameter $\Omega$, defined as $\Omega=\rho/H^2$ can
 be written as, see eg.[2],
 \begin{equation}
 \Omega=\frac{A}{A-c^2ka^{3\gamma-2} }
 \end{equation}
 If the strong energy condition is satisfied i.e. $\gamma>2/3$ then
 as the scale factor $a\rightarrow 0$ , $\Omega$ is set initially
 to 1. For increasing time $t$ the value of $\Omega$ diverges  as
 [2]
 \begin{equation}
 |\Omega-1|\propto t^{2-4/3\gamma}
 \end{equation}
 We can estimate the value of $\Omega$ at the Planck time
 and assuming $\gamma=4/3$ throughout the evolution of the universe.
 The age of the universe is $\sim 10^{60}t_{pl}$. Then using expression
 (6) we can relate $\Omega$ at different times as
 \begin{equation}
 \frac{(\Omega-1)_{now}}{(\Omega-1)_{then}}\approx 10^{60}
 \end{equation}
 If we assume that today $\Omega\approx 1$ then at the Planck
 time we require $(\Omega-1)<10^{-60}$ , i.e. $\Omega\sim
 1\pm 10^{-60}$ .
  The  flatness problem
  can be considered as being effectively  solved by having an exceedingly
  big value of $\dot{a}$ at the Planck time. This sets the density
  parameter $\Omega$, where
  \begin{equation}
  \Omega=1+\frac{c^2k}{\dot{a}^2}\;\;,
  \end{equation}
  extremely close to unity so that even today at time $\sim 10^{60}t_p$
  it has still not
  departed significantly from unity. Again the large
  value of the constant $A\sim 10^{120}$ can achieve
  this since for radiation $\dot{a}^2 =A^{1/2}t^{-1} $.

  So far we have not included any inflationary early stage
  However, with inflation the  Planck  problem
  is helped by having
  a huge expansion while the energy density remains roughly
  constant. This  obviates the
  need for arbitrary constants that usually
  set parameters, particularly $\dot {a}$,
  vastly post-Planckian where quantum gravity
  is utterly dominant. With inflation, the constant `$A$' is
  automatically forced large
  without requiring an unnaturally large initial value -see eg.[3].
  We should add that the flatness problem might not actually be a
  problem at all in SBB cosmology
  and is rather a question of how one `picks' the
  arbitrary constant `$A$'. If the equations had a different form
  or transformed to different variables then a
  large `$A$' might be quite natural. This is indeed
  the case when one considers an invariant canonical measure for
  the classical solutions [6], or works with a DeWitt
  superspace approach [7] or even with Bayesian reasoning [8].
  Likewise, as emphasized by Padmanabhan
   the horizon problem is also essentially a
  quantum gravitational problem as changing the behaviour of the scale
  factor, just while $t<t_{pl}$, can generally resolve the problem [4]. For
  these reason we regard the Planck problem as being the fundamental
  puzzle of non-inflationary (SBB) cosmology and to which
  alternative models, here the variable speed of light, must help
  resolve.

  {\bf  2.0 Variable $c$ cosmology}

  It has been suggested [9,10,11] that by changing
   the speed of light $c$ during
  the early universe the various puzzles can be solved. One is
  effectively resetting the constant $A$ above  to be unity in new
  units. Consider the Planck values
  \begin{equation}
  m_{pl}=\sqrt{\frac{\hbar c}{G}}\;\;\;:\;\;
  l_{pl}=\sqrt{\frac{\hbar G}{c^3}}\;\;\;:\;\;
  t_{pl}=\sqrt{\frac{\hbar G}{c^5}}
  \end{equation}
  Note that although the Planck mass increases with $c$, both the
  Planck size and Planck time decrease more rapidly with $c$: this will be
  shown to be the source of a fatal flaw with such alterations of $c$.

  First consider the Planck density $\sim m_{pl}/l_{pl}^3$ this scales as
  $c^5$ , so an increase in $c$ of order $10^{20}$ would appear to
  possibly resolve the Planck density problem by increasing it
  by $\sim 10^{120}$ times. It is suggested [9,10] a
  bigger increase of $10^{30}$ in $c$
  is actually needed to resolve the flatness
  problem but this actual amount will not alter our arguments.
  This higher $c$ value now allows one to apply
  the classical equations up to the
  now enormous Planck density  of $\sim 10^{240}gcm^{-3}$. We note
  in passing that the hierarchy problem of why the masses of
  elementary particles are much less than the allowed Planck mass,
  would seem exacerbated with a higher $m_{pl}$ cf.[12].

{\bf 2.1 Sudden switch in $c$}

In the first model [9,10] the speed of light was considered to undergo a
phase transition and its value to
suddenly fall by, say, a factor $\sim 10^{30}$. We will, as in ref.[10],
 represent the
first region, with higher $c$, by a  $(-)$ and the subsequent region with
$c$ taking its present value by a $(+)$, i.e. $c_+=10^{-30}c_-$.
 Now consider region $(-)$ , since
$c$ is fixed the Friedman equations remain valid. Although we do not
have a correct measure to apply at the Planck epoch in regime $(-)$ we
will assume that initially roughly equipartition is valid and that
$\Omega\simeq 1$ being given by a radiation source. This takes place
now for Planck size $l_{pl}^{-}\sim 10^{-45}*10^{-33}cm\sim 10^{-78} cm $
and Planck time $t_{pl}^{-}\sim 10^{-75}*10^{-44}$ s $\sim 10^{-119}$s.

We cannot change the value of $c$ until the scale factor and time
are greater than the Planck values in region $(+)$ otherwise the
universe is simply left stranded within the quantum gravitational
epoch when we have no realistic idea of what happens. We refer to this
as achieving Planck epoch escape.  But  this means that
the high $c$ region has to exist for $10^{75}$ of its Planck units before
the time becomes $10^{-44}$ seconds, or $t_{pl}^+$ which is also
 the present Planck time with our
value of $c$. During this period $\Omega$ will diverge away from unity
in the usual way $|\Omega-1|\propto t$ but it now has a longer time to
diverge $\sim 10^{75}$ compared  to the usual radiation big bang model
with present time $10^{60}$ Planck units.

Consider first the case of closed $(k=1)$ universes. Keeping other
constants fixed the maximum size of the closed universe $a_{max}$ is
reduced for increased $c$,
 actually $a_{max}\propto c^{-1}$ -cf. eq.(11) below. It is
even more unlikely that any closed universe will survive to time
$t_{pl}^+\sim 10^{-44}$s than a standard
 big bang model will survive to our present age. Things
are  little better for the open $(k=-1)$ cases. Because the initial
value of $\Omega$ is not initially highly tuned to be unity the curvature
will rapidly dominate the dynamics. Once the curvature dominates the
solution becomes of the Milne form and the
expansion rate increases. It is now a
faster $a\sim t$ rate
 compared to radiation $a\sim t^{1/2}$ and this contributes
 more dilution of the radiation
  matter term since $\rho\sim a^{-4}$. For example consider the
 case that $\Omega$ is initially $1-O(10^{-60})$,  the same
 required amount of
 fine tuning as in the usual big bang
  model. For the first $10^{60}t_{pl}^-$
 the scale factor grows $10^{30}l_{pl}^-$. For the remaining time
 $10^{15}t_{pl}^{-}$ until $t_{pl}^+$ the scale factor grows a further
 $10^{15}l_{pl}^-$, being driven faster in a curvature dominated phase.
 In total
 the scale factor has grown $10^{45}l_{pl}^-$ which is just equal to
 $l_{pl}^+$. When the speed of light now changes the curvature is
 diluted by a factor $c^2\sim 10^{60}$ so once more the value of $\Omega$
 is $1-O(10^{-60})$ - so no actual improvement has been made
 to the fine-tuning of
 $\Omega$. If the initial value of $\Omega$ was less fine tuned than
 $O(10^{-60})$ the curvature would have dominated earlier and the
 radiation diluted more: the value of $\Omega$ would still be roughly
 zero even after the speed of light had  changed to its lower
 value. Although this Milne curvature phase was anticipated in
 ref.[10] their analysis had not taken into account the altered Planck
 units and they missed the  extremely long period of time,
 in the Planck units of region $(-)$, that passes
 before a phase transition can
 occur. During this time the
 matter is being rapidly diluted, in total as in the above
 example, by a factor bigger than
 can be compensated by the later switch in $c$. One still needs a
 mechanism to produce matter with $\Omega\simeq 1$ at $t_{pl}^+\sim
 10^{-44}$s.

 Making the change in the speed of light even bigger would not
 help, in fact, it will make the Planck time $t_{pl}^-$ even
 smaller and allow even more time for $\Omega$ to depart from
 unity. Working with a dust equation of state $(\gamma=1)$ gives,
 allowing for changes in expansion rate , a similar result. It
 might be thought that decreasing $\gamma$ would eventually allow
 one to succeed, but recall the density now decreases slower $\rho \propto
 a^{-3\gamma}$. One must wait until the initial density $\sim 10^
 {240}gcm^{-3}$ falls below $\sim 10^{93}gcm^{-3}$ before one can
 change the value of $c$, cf. eq.(2)  above.

 To conclude, the flatness problem is either
 worsened or just remains the same depending
 on the initial degree of fine tuning. We note in passing that a
 matter source with `curvature' equation of state
 $\gamma=2/3$ would not be diluted,  and
 the change in $c$ could indeed set the value of $\Omega=1\pm O(10^{-60})$.
 But this, so-called `coasting' solution
 case is already known to be borderline inflationary [2].

  As for the Planck
 problem, it has, essentially by fiat,
 been solved by redefining the large constant `$A$' to unity but
 then we need to understand why $c$ then suddenly
 changes later in the universe's
 evolution. This has to occur at a time huge in
 the Planck time units of this high $c$
 universe when `quantum gravitational' effects are then not expected to
 dominate. Why $c$ should change simultaneously over such scales is also
 unclear. Since the transition proceeds rapidly the regions are rapidly
 loosing causal contact and then why they should all choose the same $c_+$
 seems a further complication.
  The Planck problem has just been rewritten in a new guise which
 is now just as arbitrary and unexplained: previously we did not
 understand why `$A$' was large, then constants are set to make it
 appear natural with the value
 unity, but this then requires that $c$
  change by a huge factor $\sim10^{30}$ much  later in the evolution of
  the universe: again the ubiquitous  mismatch of scales.

 The horizon problem does not seem to be explained as well as by
 using inflation. The $(-)$
 region has its own horizon problem
  as $t\rightarrow 0$ ; $ \dot{a}\rightarrow
 \infty$: so the expansion rate can always `outdo'  $c$. Any model
 with an expansion $a\sim t^p$ with $0<p<1$ will suffer this
 divergence in $\dot{a}$ as $t\rightarrow 0$. Unlike inflation
 one cannot get the universe emanating from a single region that has
 always being in causal contact as $t\rightarrow 0$.
  Why the horizon problem is then `solved' just because it
 exists for a time $10^{75}t_{pl}^-$ makes no more sense than
 saying the horizon problem would be
  presently solved in any SBB model that
 grew to our present universe. Some
 mechanism is required to explain this smoothing and also the fact
 that fluctuations that can grow during this $10^{75}t_{pl}^-$
 period will need to be sufficiently erased. Arguments based on
 perturbation theory will hardly suffice given such large time scales for
 evolution cf.[9,10]. Simply claiming the Jeans mass
 is never reached during the
 high $c$ period, so that no
 structures can form, seems simply wishful thinking cf.[10].
   There is also a  dubious argument [9] to give
 a scale invariant spectrum; but this uses the $(+)$ Planck values in
 the $(-)$ region and anyway uses an inflationary result: the presence of
 Hawking radiation, for the
 generation of fluctuations. Also note that
  any cosmological
 constant will  have ample time to become dominant before $c$
 switches unless $\Lambda$, for some reason, is
  already extremely small in the
 $(-)$ region.

 To summarize the horizon
 problem: although
 superficially it appears that the horizon $\sim ct$ can be
 resolved by a big increase in $c$ the natural time unit is
 correspondingly reduced at a faster rate $\propto c^{-5/2}$. The Planck
 horizon size $\sim ct$ correspondingly
  falls from $10^{-33}$cm to $10^{-79}$cm. This
 means that the universe has to exist for huge times to create a
 sufficiently large causal region,
  but even then no `smoothing mechanism' is
 presented. If one postulates such a mechanism  then
  why doesn't this mechanism still keep the universe smooth
 today as it is also `only' at age $10^{60}t_{pl}^{+}$?
 Essentially this is just another way of saying
 that one can always re-set units so that  $c=1$
  in region $(-)$ and no extra phenomena  is really being introduced.

To conclude this section: little advantage has been found by
invoking a sudden change in the speed of light. The flatness
problem remains and the Planck problem is just transformed as to:
why the speed of light should change by an enormous amount
 when the universe is hugely
larger than its natural Planck units? Instead we next consider the
alternative and more extreme contention that $c$ is continually changing.

{\bf 2.2 Gradual changing of $c$}

 It is also possible that the speed of light changes gradually instead
 of being a sudden jump. Although we leave aside worries that this
 would seem to contradict various experimental data dating back to
 shortly after the big bang: the $c$
 changers would contend that other variables would also change to
 compensate. Consider now the relevant equations [10,11,13].
 \begin{equation}
 H^2+\frac{c^2(t) k}{a^2}=\rho
 \end{equation}
 \begin{equation}
 \dot{\rho} +3H\left ( \rho+\frac{p}{c^2(t)}\right )=\frac{2kc\dot{c}}{a^2}
 \end{equation}
  The matter obeying the usual equation of state
  \begin{equation}
  p=(\gamma-1)\rho c^2(t)
  \end{equation}
  Taking the speed of light to alter with the scale factor, such that
  \begin{equation}
  c=c_o a^n
  \end{equation}
  with $c_o>0$ and $n$ constants.

  Because there are no longer two distinct regions where Planck constants
  can be fixed the analysis is
  more involved but is constrained  in a similar
  fashion. Recall that we need
  to ensure that the scale factor remains larger than the Planck length,
  and of course that the age of the universe never be less than the current
  Planck time. At the same time we need to be diluting the curvature more
  rapidly than the fall off in matter density to ensure $\Omega$ stays
  near unity. Solving the above equations it is found that [13]
  \begin{equation}
  \frac{\Omega}{\Omega-1}= Ba^{2-2n-3\gamma} +
  \frac{C}{2n-2+3\gamma}\;\;\;,
  \end{equation}
   with $A$ and $B$ constants. To give flatness ($\Omega\rightarrow 1$)
   as $a\rightarrow \infty$ requires
  $n<(2-3\gamma)/2$ where the expansion asymptotes to the usual
  $a\sim t^{2/3\gamma}$ behaviour [11,13,14]. The same bound can also solve
  the horizon problem [11,13].

  We concentrate on the Planck escape aspect which is first necessary
   to resolve but which will be found to then constrain whether the
   flatness and horizon problems can also be simultaneously solved.
   For an expansion  $a\sim t^{2/3\gamma}$
   the time goes as $t\propto a^{3\gamma/2}$. This should be
   contrasted with the Planck time $t_{pl}$ which scales as $\sim
   c^{-5/2}$, or using the relation above $t_{pl}\propto
   a^{-5n/2}$. Since we require that $t>t_{pl}$ for increasing $a$
    to stay away from the
   quantum gravitational epoch we get a constraint on the allowed
   negativity of $n$, such that $n>-3\gamma/5$. This should now be
   contrasted with the required value that was obtained to resolve
   the flatness problem $n<(2-3\gamma)/2$. These two constraints
    are now only compatible
       for $0\leq \gamma< 10/9$,
    so excluding the important radiation case $\gamma=4/3$. One has
    extended the inflationary producing value of $0\leq\gamma\leq 2/3$ that
    always resolves the flatness and horizon problems  just up
    to $10/9$. Although this now includes, unlike the
    sudden change in $c$ example, the dust $(\gamma=1)$ case
     this seems  a large price to pay for such an
    advantage. I also leave aside doubts that in resolving
    the horizon problem for $2/3<\gamma<10/9$,
    one is still taking the variables `out of bounds' into
    the quantum gravitational
    regime.

     Now it is then further argued [11,13,14] that the increasing speed
    of light does also have  the further
     advantage, over inflation, of resolving the
    cosmological constant problem: why $\Lambda \simeq 0$.
     But this requires an even  larger
    negative $n$ such that $n<-3\gamma/2$ [11,13] which is again not
    compatible with the Planck epoch escape requirement for any
    value of $\gamma$.

    Alternatively
    for $0>n>-3\gamma/2$ it was claimed [11,13] that, so-called
     quasi-flatness could be
    achieved: where the ratio of matter density to cosmological constant
    approaches a constant. But for this the scale factor then goes as
    $a\propto t^{-1/n}$ [11,13]. This means $t\propto a^{-n}$ and
    comparing with the Planck time $t_{pl}\propto a^{-5n/2}$ gives
    $t/t_{pl}\propto a^{3n/2}$. So this behaviour of the scale factor
    would now require $n>0$ to escape
    the Planck epoch which is in contradiction with  the
    solution i.e. $n<0$, that would give quasi-flatness. The necessity of
    Planck epoch escape makes it no longer possible to have a
    sufficiently changing speed of light to set $\Lambda\rightarrow 0$
    or even one that could  achieve quasi-flatness. On would need to go
    outside the range of validity of variable $c$
    General relativity theory in
    the attempt to solve the cosmological constant problem. In
    other words one is actually working within
    the unknown Planck domain in
    order to solve the various problems. It is known, without even
    changing $c$, that simply ignoring the Planck epoch
    circumvents the flatness and Planck problems: essentially
    because all FRW universes are equivalent modulo arbitrary
    constants cf.[6-8]. In an apparent response to these sorts of
    criticism Barrow and Magueijo [15] try to defend the constantly
    changing $c$ theory. For the case of radiation they give the
    expression $t_{pl}\propto t^{-n}$.
    \footnote{ I seem to obtain the
    relation $t_{pl}\propto t^{-5n/4}$ but this discrepancy does
    not alter the following arguments.}

     To solve the various problems
    requires $n<-1$, for the sake of argument take $n=-2$, then
    $t_{pl}\propto t^2$. In ref.[15] they note that as $t\rightarrow
    0$, $t_{pl}$ approaches zero more rapidly
     than $t$ so that in some sense the
    Planck time is never approached. But I have rather emphasized
    the problem that occurs for $t>>1$ when $t_{pl}/t\propto t$.
    So it is more `natural' for $t_{pl}$ to be greater than $t$
    unless arbitrary large constants are imposed to prevent this.
    You might try and argue that we never reach cosmic time $t>1$ by using
    some large time units, but this would simply introduce another
    problem of why there are different time scales. \footnote{ As mentioned
    in a
    purely classical model there is no natural time scale or `ruler'
     to judge scales by, but  in
     practice there are natural length scales cf. [8] and such
     problems of scale cannot be transformed away.}

    In the limit $n\rightarrow -1$ we obtain the relation $t_{pl}
    \propto t$ which is still unsatisfactory since at present
    $t/t_{pl}\sim 10^{60}$ and we wish to understand this large
    number.

    The decreasing value of $c$ is making the Planck values rapidly grow
    and eventually overtake the regions of classical validity.
    In some sense the universe starts out in a classical domain
    for $t\rightarrow 0$ and then becomes quantum gravitational
    dominated
    as $t>1$. This
    is the opposite of how the big bang is usually perceived in
    that quantum gravity is assumed necessary as $t\rightarrow 0$.
   The advocates have assured me that ``there is no principle
   stating that the universe must be created in the Planck
   epoch'', but this does not seem a virtue.

    The lack of an initial quantum gravitational epoch  means the theory is
    never superseded and the various expressions and
    initial conditions have to be
    simply taken as given without any further explanation. A
    quantum gravitational epoch
    might also have helped describe or ameliorate the still
    present
    singularity at $a\rightarrow 0$. In any case there is still a,
    now apparently redundant, quantum gravitationally
    scale in the universe that
     becomes increasingly important in the future ( actually
     for when $t>1$) as $c$ decreases. They say that ``instabilities
     of the big bang are converted into attractors by varying c'':
     but it seems that along with this the quantum epoch is also
     switched, to occur at now future times.

    To conclude, the only advantage we find with a constantly changing $c$
    is that there is a
    slight improvement in the resolution of the
    flatness and horizon problems
    from the usual strong energy violating or inflationary
    values $0\leq\gamma\leq2/3$
     to now $0\leq\gamma<10/9$. This seems a high price to pay for
     such little improvement  especially when
     other possible problems could further restrain how $c$ may vary, which
     would decide what values of
     $n$ are allowed. This advantage is further offset by then needing to
     explain why $c$, although  initially tending to $\infty$ at
     the initial singularity, changes at some unexplained rate.
     Incidentally this $\infty$ being the reason why the horizon problem
     can be solved compared to the previous finite sudden change in $c$.
     The initial singularity is also effectively null in this
     case for infinite $c$.

      We have not allowed other
     constants to vary, particularly Newton's constant $G$, and
     this might allow further scope to give a natural explanation
     of SBB cosmology. However, we remain rather skeptical. A
     related attempt to use Brans-Dicke gravity ( so $G$ can vary)
      in the, so-called,
     pre-big bang cosmology also suffers a related Planck problem
     [16]. This is because the strong-energy condition is
     likewise not being
     violated and it is then not clear why the universe is being driven
     large unless arbitrary constants cf. `$A$'
      are again picked to be huge. Switching
     around constants seems more an exercise in
      `rearranging the deck chairs' while what seems to be  of more
       fundamental use is actually making gravity repulsive:
       inflation.

       This might be somewhat unfair as allowing changing
       `constants' might have other more aesthetic advantages. But
       these advantages need to be carefully assessed and placed
       alongside the disadvantage of requiring
       somewhat ad hoc assumptions of how these `constants' should
       change. To avoid simply rewriting puzzles
        in new ways, one needs a theory that gives a certain prediction
       for how these changes should occur.

       Recall, that anyway one can simulate
       some of the possible  advantages of
       higher $c$ values without actually altering $c$ or any of the
       usual fundamental constants:  examples include
       allowing wormholes during the early universe [17] or simply
       having more extreme geometries
       with closed timelike curves [18].

\newpage

{\bf References}\\
\begin{enumerate}
\item Y.B. Zeldovich, ``My Universe: selected reviews'', Harwood Academic
Press (1992) p.95.
\item G.F.R Ellis, ``Proc. Banff Summer Research Institute on
Gravitation'' eds. R. Mann and P. Wesson (World Scientific:
Singapore) (1991).\\Class. Quant. Grav. 5 (1988) p. 891.
\item A.D. Linde, ``Particle Physics and Inflationary cosmology'',
(Harwood: Switzerland) (1990).\\
E.W. Kolb and M.S. Turner, ``The Early Universe'' (Addison-Wesley:New
York) (1990).
\item T. Padmanabhan, {\em Structure formation in the universe}
Cambridge University press: Cambridge (1993) p.359.
\item J.M.M. Senovilla, Gen. Rel. Grav. 30 (1998) p.701.
\item S.W. Hawking and D.N. Page, Nucl. Phys. B 298 (1988)
p.789.\\
D.H. Coule, Class. Quant. Grav. 12 (1995) p.455.
\item H.T. Cho and R. Kantowski, Phys. Rev. D 50 (1994) p.6144.
\item G. Evrard and P. Coles, Class. Quant. Grav. 12 (1995)
p.L93.\\
D.H. Coule, Class. Quant. Grav. 13 (1996) p.2029.
\item J.W. Moffat, Int. J. of Mod. Phys. A 2 (1993) p.351.
\item A. Albrecht and J. Magueijo, Phys. Rev. D 59 p.043516.
\item J.D. Barrow,  Phys. Rev. D 59 (1999) p.043515.
\item A.D. Sakharov, Sov. Phys. Doklady 12 (1968) p.1040.
\item J.D. Barrow and J. Magueijo, ``Solutions to
the Quasi-Flatness and Quasi-Lambda problems'', preprint astro-ph/9811073.
\item J.D. Barrow and J. Magueijo, ``Varying-$\alpha$ Theories and
solutions to the cosmological problems'', preprint
astro-ph/9811072.
\item J.D. Barrow and J. Magueijo, `` Solving the flatness and
quasi-flatness problems in Brans-Dicke cosmologies with a varying light
speed'',preprint astro-ph/9901049.
\item D.H. Coule, Class. Quant. Grav. 15 (1998) p.2803;\\
 Phys. Lett. B 450 (1999) p.48.
\item L.Liu, F. Zhao and Li-Xin Li, Phys. Rev. D 52 (1995) p.4752.
\item J.R. Gott and Li-Xin Li, Phys. Rev. D 58 (1998) p. 023501.
\end{enumerate}
\end{document}